\documentclass[showpacs,amsmath,reprint,showkeys,pra]{revtex4-1}
\usepackage{dcolumn}    
\usepackage{bm}         
\usepackage{ifpdf}
\usepackage{amssymb,lineno,amsfonts}
\usepackage{graphicx}   
\usepackage{bbm}
\usepackage{mathrsfs}
\usepackage{upgreek}
\usepackage{epstopdf}
\usepackage{setspace}
\usepackage{hyperref}
\usepackage[matrix,frame,arrow]{xypic}
\usepackage{float}
\usepackage{soul}
\usepackage{tikz}
\usepackage{natbib}
\usepackage{color} 
\usepackage{subfigure}
\usepackage{tikz}
\usepackage[]{qcircuit}

\definecolor{med-blue}{RGB}{25,25,112} 
\hypersetup{colorlinks, linkcolor={blue},citecolor={blue}, urlcolor={black}}
\newcommand{\ket}[1]{\vert{#1}\rangle}

\newcommand{\outpr}[2]{\vert{#1}\rangle\langle{#2}\vert}

\newcommand{\expv}[1]{\langle{#1}\rangle}

\newcommand{\proj}[1]{\outpr{#1}{#1}}

\newcommand\at[2]{\left.#1\right|_{#2}}
\newcommand{\Arrow}[1]{%
\parbox{#1}{\tikz{\draw[->](0,0)--(#1,0);}}
}

\begin{document}
	\title{Star-topology Registers: NMR and Quantum Information Perspectives}
	\author{T S Mahesh}
	\email{mahesh.ts@iiserpune.ac.in}
	\author{Deepak Khurana}
	\email{Current Affiliation: Section of Quantum Physics and Information Technology (QPIT), Department of Physics, Technical University of Denmark, 2800 Kgs. Lyngby, Denmark. Email: deekh@dtu.dk. }
	\author{Krithika V R}
	\email{krithika\_vr@students.iiserpune.ac.in}
	\author{Sreejith G J}
	\email{sreejith@acads.iiserpune.ac.in}
	\author{C S Sudheer Kumar}
	\email{sudheer.kumar@students.iiserpune.ac.in}
	\affiliation{Department of Physics and NMR Research Center,\\
		Indian Institute of Science Education and Research, Pune 411008, India}
	
\begin{abstract}
{Quantum control of large spin registers is crucial for many applications ranging from spectroscopy to quantum information. A key factor that determines the efficiency of a register for implementing a given information processing task is its network topology.  One particular type, called star-topology, involves a central qubit uniformly interacting with a set of ancillary qubits.  
A particular advantage of the star-topology quantum registers is in the efficient preparation of large entangled states, called NOON states, and their generalized variants.  
Thanks to the robust generation of such correlated states, spectral simplicity, ease of polarization transfer from ancillary qubits to the central qubit, as well as the availability of large spin-clusters,  the
 star-topology registers have been utilized for several interesting applications over the last few years. 
Here we review some recent progress with the star-topology registers, particularly via nuclear magnetic resonance methods. 
}
\end{abstract}

\keywords{}
\maketitle	
	
\section{Introduction}
The long-lasting quantum memory of nuclear spins is at the heart of versatile applications of nuclear magnetic resonance from spectroscopy and biomedical imaging to quantum information processing.  
While remarkable progress has been achieved with few-spin systems forming small quantum registers, scaling the register size has been a daunting task.  The challenges include the highly mixed state of NMR ensembles, spectral complexity of large registers that limit their quantum control, and decoherence causing the loss of quantum memory.  In this review, we explain how certain symmetries, particularly the star-symmetry, can provide a way forward to realize large quantum registers, albeit for specific applications.

\begin{figure}
	\centering
	\includegraphics[trim=0cm 11cm 0cm 0cm,width=7cm,clip=]{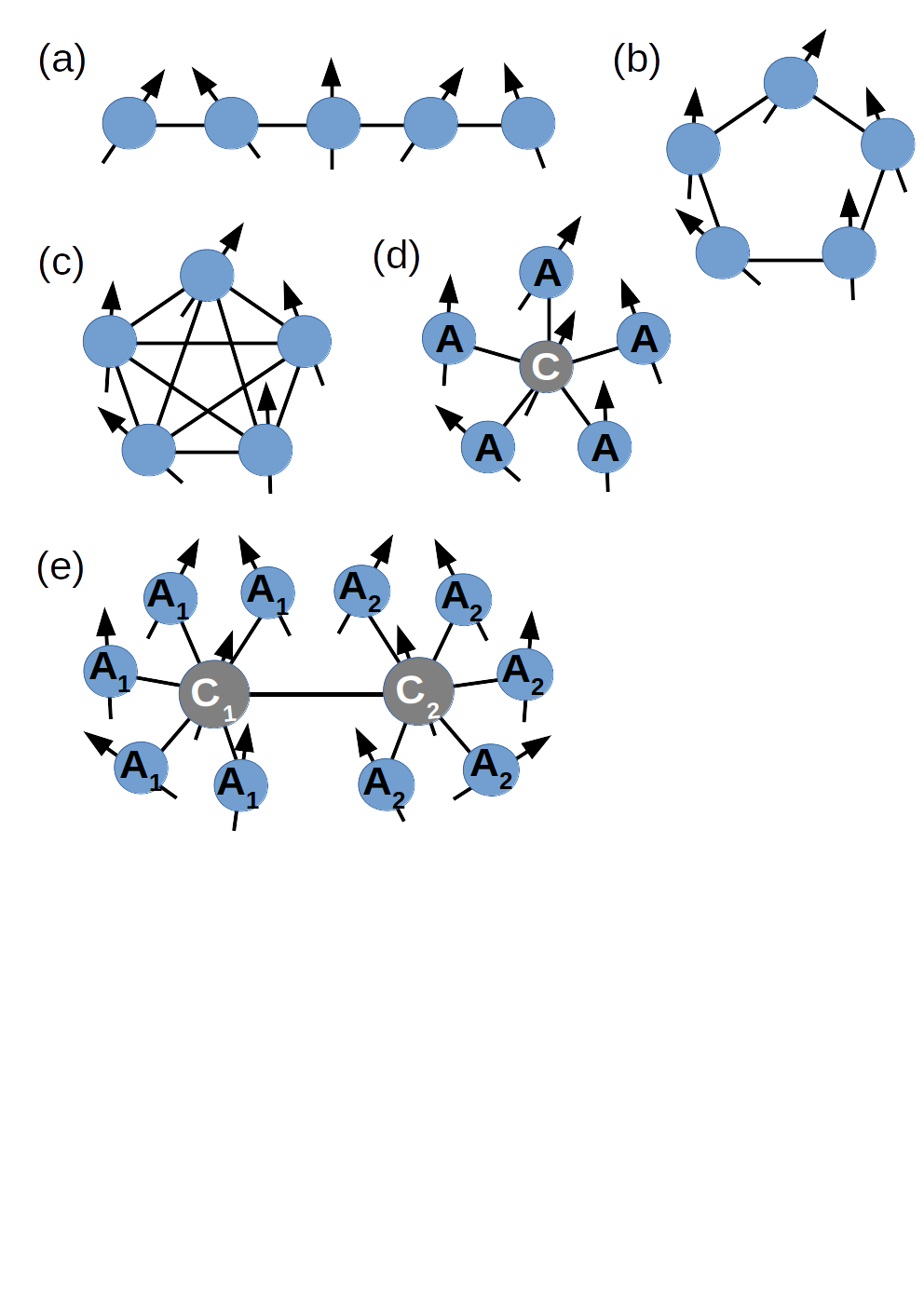}
	\caption{Some common topologies of quantum registers: (a) linear chain, (b) cyclic chain, (c) all-to-all, (d) central-spin or star-topology register (STR), and (e) double-STR.}
	\label{tops}
\end{figure}

Quantum registers may be categorized based on the network-topology of qubits.  Some common topologies are shown in Fig. \ref{tops}.  In principle, linear and cyclic chains are sufficient to realize universal quantum gates, while all to all topology is most efficient, since it allows a direct transport of information from one qubit to any other.  As we scale up the size of the register though, it is hard to maintain all to all interactions, and it will become necessary to consider a practical topology that is best suited for a particular application.
In this review, we focus on the star-topology registers (STRs), which consist of a central qubit uniformly interacting with a set of ancillary qubits.  We shall discuss various aspects of star-topology registers, mainly from the perspectives of nuclear magnetic resonance (NMR) and quantum information. Although here we treat a qubit synonymous with a spin-1/2 particle, the underlying principles hold in non-NMR architectures.

For completeness, we shall first introduce the theoretical and spectroscopic aspects of the STRs in the following section. In section III, we shall review recent progress with utilizing the STRs for various applications.  Finally, we shall summarize in section IV.

\section{Star-topology register (STR)}
\label{STR}
As illustrated in Fig. \ref{tops}(d), an STR consists of a central qubit (C) directly connected to a set of ancillary (A) or satellite qubits.  Many of the STRs realized by nuclear spin-systems via liquid-state NMR techniques have the following properties:
(i) no net effective interactions among the ancillary qubits;
(ii) the central qubit uniformly interacts with each of the ancillary qubits;
(iii) the central qubit is realized by a different nuclear isotope and hence can be selectively addressed;
(iv) the ancillary qubits are indistinguishable from one another.  
The last point implies that the ancillary qubits can only be collectively addressed, but no selective addressing of a particular ancillary qubit is possible.  While this fact limits the type of quantum gates that can be realized, crucially, it also brings about spectral simplicity that allows us to conveniently handle much larger registers.

We shall consider an STR with $N-1$ ancillary qubits surrounding a central qubit.  For such a system, the NMR Hamiltonian (in units of $\hbar$)
can be written as 
\cite{levitt2013spin}
\begin{eqnarray}
H_0 &=& H_C + H_A + H_{CA}
\nonumber \\
&=& \omega_{C} I_z^C +\omega_{A} I_z^A + 2\pi J_{CA} I_z^CI_z^A,
\end{eqnarray}
where $\omega_{C}$ and $\omega_{A}$ 
are the Larmor frequencies of $C$ and $A$ qubits,
$J_{CA}$ is the fixed strength of the indirect spin-spin interaction that is characteristic of the spin-system, $I_z^C$ is the spin operator for the central qubit and $I_z^A = \sum_{k=1}^{N-1} I_z^{A,k}$ is the collective spin operator for the ancillary qubits.  
The Larmor frequencies are proportional to the strength of the magnetic field $B_0$,
\begin{eqnarray}
\omega_C = -\gamma_C B_0 ~~\mbox{and}~~ \omega_A = -\gamma_A B_0,
\end{eqnarray}
where $\gamma_C$ and $\gamma_A$ are the gyromagnetic ratios of the nuclei.

\begin{figure}
	\centering
	\includegraphics[trim=0cm 17cm 0cm 1cm,width=9cm,clip=]{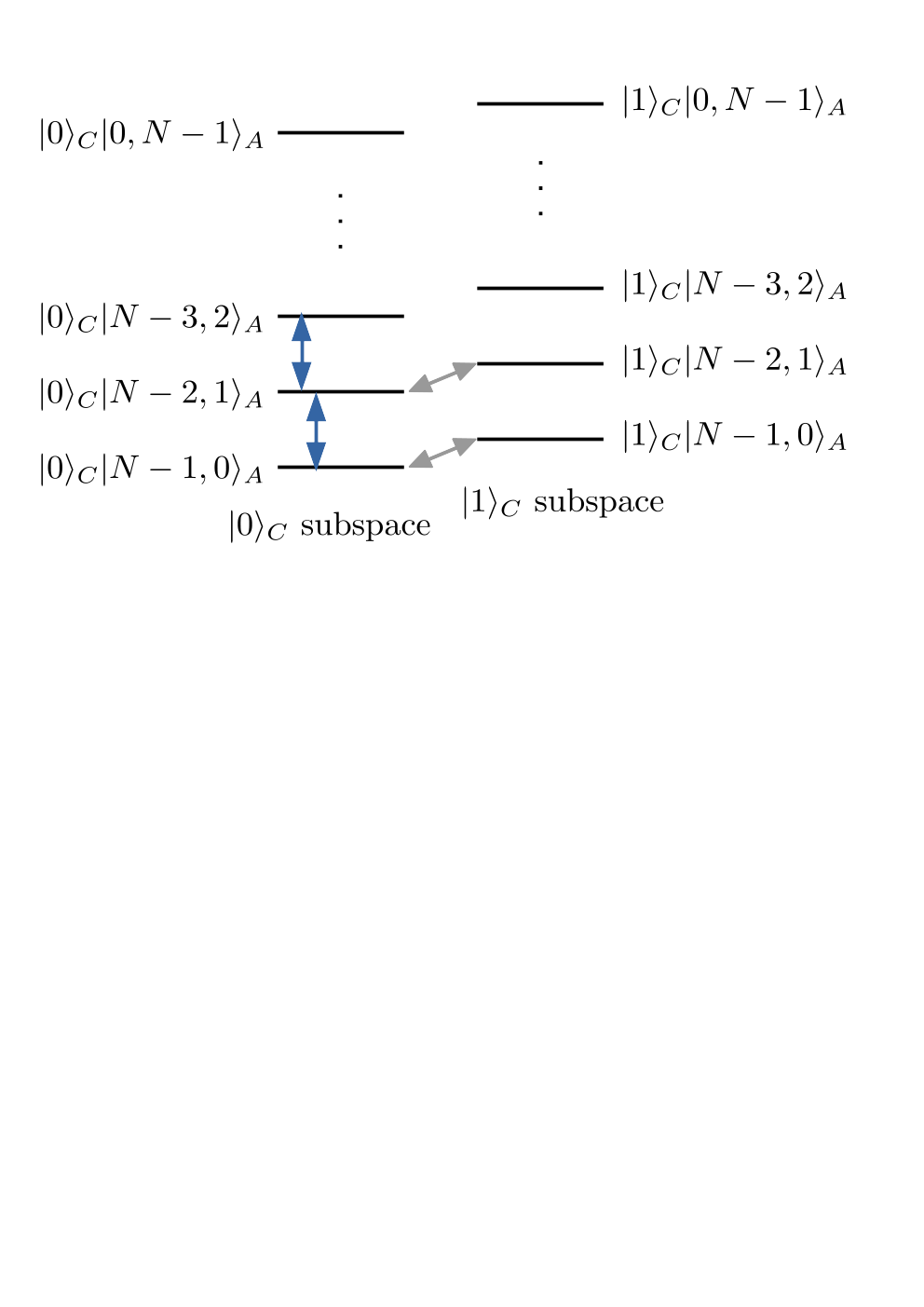}
	\caption{Energy levels of an STR forming two subspaces corresponding to the $\ket{0}_C$ and $\ket{1}_C$ states of the central spin.  The transitions of $A$ and $C$ are indicated by blue and gray arrows respectively.}
	\label{starlevels}
\end{figure}

The eigenstates of STR are Zeeman product states of individual qubit-states $\ket{\uparrow_z} \equiv \ket{0}$ and $\ket{\downarrow_z} = \ket{1}$.
Thus, we label the STR eigenstates with $N$-bit binary strings and form a computational basis.  Each binary string is associated with a Hamming weight $\mathbbm{h}$ that counts the number of 1's in the string.  The labeling is chosen such that ascending energy levels correspond to ascending Hamming weights and degenerate levels get the same Hamming weight.  
Levels of same Hamming weight are collectively represented by $\ket{N-\mathbbm{h},\mathbbm{h}}$.
Similar labeling schemes had been conveniently employed for information encoding purposes even when the eigenstates  are not product states \cite{sinha2001toward,mahesh2002ensemble,murali2002quantum}.
The total spin magnetic moment quantum number for the state vector $\ket{N-\mathbbm{h},\mathbbm{h}}$ is thus
\begin{eqnarray}
m_{\mathbbm{h}} = 
\frac{N}{2} - \mathbbm{h}.
\end{eqnarray}
Similarly, if ${\tt h}$ is the Hamming weight for the ancillary part of the state vector, i.e., $\ket{N-1-{\tt h},{\tt h}}_A$, then
\begin{eqnarray}
m_{{\tt h}}^A = \frac{N-1}{2} - {\tt h},
\end{eqnarray}
is the total ancillary spin magnetic moment quantum number.
As illustrated in Fig. \ref{starlevels}, the energy eigenstates form two subspaces corresponding to $\ket{0}_C$ and $\ket{1}_C$ states of the central qubit: \\

\noindent
\begin{tabular}{p{3.1cm}|p{2.7cm}|p{2.5cm}}
\hline 
Eigenstates \hspace{1cm}
${\tt h}\in\{0,1,\cdots,N-1\}$ & Total magnetic quantum number & Eigenvalue \\
\hline 
$\ket{0}_C\ket{N-1-{\tt h},{\tt h}}_A$ &
$m_{0,{\tt h}} = \frac{1}{2}+m_{{\tt h}}^A$ &
$+\omega_{C}/2+m_{{\tt h}}^A(\omega_{A}+\pi J_{CA})$ \\
\hline
$\ket{1}_C\ket{N-1-{\tt h},{\tt h}}_A$ &
$m_{1,{\tt h}} = -\frac{1}{2}+m_{{\tt h}}^A$ &
$-\omega_{C}/2+m_{{\tt h}}^A(\omega_{A}-\pi J_{CA})$ \\
\hline
\end{tabular} \\

\begin{figure}
	\centering
	\includegraphics[trim=0cm 0cm 0cm 0cm,width=7cm,clip=]{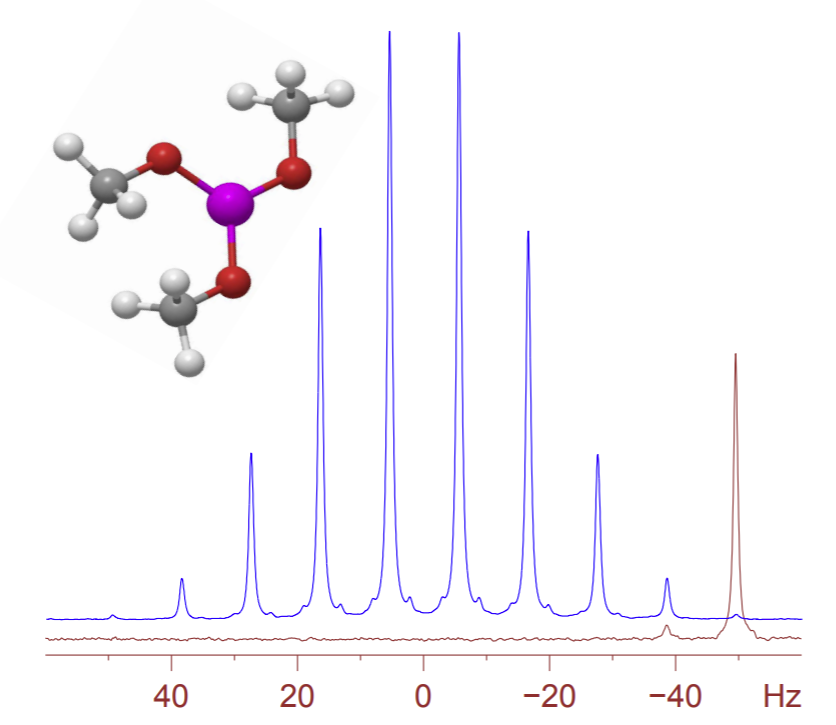}
	\caption{$^{31}$P spectra of trimethylphosphite corresponding to single quantum excitation from thermal equilibrium (upper trace) and corresponding to the NOON state (lower trace). The molecular structure of trimethylphosphite shows the central $^{31}$P spin (in pink) that is coupled to nine $^1$H spins (in white) via an indirect scalar coupling of $J_{CA} = 11$ Hz.}
	\label{tmpspec}
\end{figure}

The selection rule for the radio frequency (RF) excitation allows only transitions with a unit change in the total spin magnetic moment quantum number. Thus, all inter-subspace transitions of $C$-spin are allowed, while $A$-spin transitions within each subspace satisfy $\Delta m_{\tt h}^A = -\Delta {\tt h} = \pm 1$.
As a result, C-qubit possesses $N$ distinct transition frequencies  $\omega_C+2\pi J_{CA}m_{\tt h}^A$ (gray arrows in Fig. \ref{starlevels}).  On the other hand, A-qubits collectively have only two distinct transition frequencies, $\omega_A + \pi J_{CA}$ in the $\ket{0}_C$ subspace and $\omega_A - \pi J_{CA}$ in the $\ket{1}_C$ subspace.  Fig. \ref{tmpspec} shows the experimental spectrum of $^{31}$P spin coupled with nine ancillary $^1$H spins in trimethyl phosphite.


Owing to mathematical ease, most of the NMR experiments are described in a rotating frame, that precesses with the applied RF drive, thereby rendering it time-independent. 
Since $C$ and $A$ are of different nuclear isotopes, their Larmor frequencies differ over several tens or even hundreds of megahertz.  This allows them to be controlled by independent radio-wave channels.  In a doubly rotating frame, precessing at the respective carrier frequencies, the Hamiltonian is of the form,
\begin{eqnarray}
H &=& -2\pi\nu_C I_z^C - 2\pi\nu_A I_z^A 
\nonumber \\
&&+2\pi J_{CA} I_z^CI_z^A + \Omega_C (I_x^C \cos \phi_C +I_y^C \sin \phi_C )
\nonumber \\
&&+ \Omega_A (I_x^A \cos \phi_A +I_y^A \sin \phi_A ).
\end{eqnarray}
Here $\nu_C,~\nu_A$ are resonance offsets, $\Omega_C,~ \Omega_A$  are RF amplitudes on the respective channels, and $\phi_C,~\phi_A$ are the RF phases, all of which being tunable experimental parameters.  

\subsection{Thermal state with low purity}
Under normal NMR conditions, thermal energy is much higher than the Zeeman energy gaps, so that for a given spin-1/2 nucleus,
\begin{eqnarray}
\rho = \frac{1}{2}e^{-H/kT} \approx \frac{1}{2}\mathbbm{1}+\epsilon I_z
\end{eqnarray}
where 
$\epsilon = \hbar\gamma B_0/(kT) \ll 1$ is known as the purity factor, and $\mathbbm{1}$ is the $2\times 2$ identity matrix.
Defining $\epsilon_C = \hbar\gamma_C B_0/(kT)$ and $\epsilon_A = \hbar\gamma_A B_0/(kT)$, we obtain
	\begin{eqnarray}
	\rho_C &=& \frac{1+\epsilon_C}{2}\proj{0} + \frac{1-\epsilon_C}{2}\proj{1}
	\nonumber \\
	\rho_A &=& \left[\frac{1+\epsilon_A}{2}\proj{0} + \frac{1-\epsilon_A}{2}\proj{1}\right]^{\otimes(N-1)}, ~~\mbox{so that}
	 \nonumber \\
	\rho_{CA}  &=& \rho_C \otimes \rho_A  
	\nonumber \\
	&&\approx 
	\sum_{{\tt h}=0}^{N-1} p_{0,{\tt h}} \rho_{0,{\tt h}}
	+
	 p_{1,{\tt h}} \rho_{1,{\tt h}}.
	\label{starthermalstate}
	\end{eqnarray}
	Here $\rho_{0,{\tt h}}$ corresponds to an uniform mixture of degenerate levels  $\ket{0}_C\ket{N-1-{\tt h},{\tt h}}$ of $\ket{0}_C$ subspace, while $\rho_{1,{\tt h}}$ corresponds to that of $\ket{1}_C\ket{N-1-{\tt h},{\tt h}}$ of $\ket{1}_C$ subspace.
    At low purity, i.e., $\{\epsilon_A,\epsilon_C\}\ll 1$, the relative populations are
	\begin{eqnarray}
	p_{0,{\tt h}} 
	&\approx& 
{{N-1}\choose{{\tt h}}}\frac{1+\epsilon_C+2m_{{\tt h}}^A\epsilon_A}{2^N}~~ \mbox{and,}
\nonumber \\
	p_{1,{\tt h}} 
&\approx& 
{{N-1}\choose{{\tt h}}}\frac{1-\epsilon_C+2m_{{\tt h}}^A\epsilon_A}{2^N},
	\end{eqnarray}
where ${{N-1}\choose{{\tt h}}}=\frac{(N-1)!}{(N-1-{\tt h})!{\tt h}!}$ denotes the binomial coefficient.

Thus the thermal magnetizations are  \begin{eqnarray}
M_{C} &=& \sum_{{\tt h}=0}^{N-1} (p_{0,{\tt h}}-p_{1,{\tt h}})/2
= \frac{\epsilon_C}{2},
~\mbox{for}~ C~ \mbox{spin, and}
\nonumber \\
M_{A} &=& \sum_{{\tt h}=0}^{N-1} (p_{0,{\tt h}}+p_{1,{\tt h}})m_{{\tt h}}^A
= (N-1)\frac{\epsilon_A}{2} ~\mbox{for}~ A~ \mbox{spins}.~~~~~
\end{eqnarray}
Later, we shall see how we can exploit the large reserve magnetization present in the ancillary qubits.

\section{Preparing many-body entanglement: NOON and MSSM states}
One of the key advantages of STRs is the efficiency with which the central qubit can get correlated with the ancillary qubits. Since the central qubit is uniformly coupled to all the indistinguishable ancillary qubits, a single CNOT gate can simultaneously act on all the ancillary qubits, thus offering the most efficient way to prepare many-body entanglement.
We shall consider the following cases.

\subsection{Pure ground state}
In this case, each of qubits is initially prepared in the ground state, i.e., $\ket{\psi_C} = \ket{0}$ and $\ket{\psi_A} = \ket{0\cdots 0} = \ket{N-1,0}$.  Now consider the following circuit.
\\
\hspace*{3cm}
\Qcircuit @C=1em @R=0.6em {
\lstick{\ket{\psi_C}}& \gate{H} & \ctrl{1} & \qw &
	\push{\rule{-1em}{2em}\equiv}
	& \multigate{1}{\{q_{\tt h}\}} & \qw  \\
\lstick{\ket{\psi_A}}	&  \qw {/} & \targ & \qw & & \ghost{{\{q_{\tt h}\}}} & \qw
} \\
\\

Here the Hadamard gate $H$ takes $\ket{0}$ to the uniform superposition $(\ket{0}+\ket{1})/\sqrt{2}$, while the CNOT gate, implemented simultaneously on all the ancillary qubits of the STR, flips the ancillary qubits if the central qubit is in $\ket{1}$ state, else it does nothing.
The circuit works as follows \cite{nielsen00}.
	\begin{eqnarray}
&&\ket{0}_C\ket{N-1,0}_A
\stackrel{H^C}{\longrightarrow}
\nonumber  \\
&&\frac{1}{\sqrt{2}}\ket{0}_C\ket{N-1,0}_A+
\frac{1}{\sqrt{2}}\ket{1}_C\ket{N-1,0}_A
\stackrel{CNOT}{\longrightarrow}
\nonumber  \\
&&\frac{1}{\sqrt{2}}\ket{0}_C\ket{N-1,0}_A+
\frac{1}{\sqrt{2}}\ket{1}_C\ket{0,N-1}_A
\nonumber \\	
&&=
\frac{\ket{N,0}+\ket{0,N}}{\sqrt{2}}
=\ket{\mathrm{NOON}}.
	\end{eqnarray}
The output state is named this way because of $N,0,0,N$ appearing in the state vector representation.  It is a maximally entangled state popularly known as the \textit{cat} state.  
	
\textit{Coherence order ($q_{\tt h}$):}		
For a superposition of two basis states, we define the order of coherence $q_{\tt h}$ as the difference in the Hamming weights of the two state vectors.  For an STR prepared in the superposition $\ket{0}_C\ket{N-1-{\tt h},{\tt h}}_A+\ket{1}_C\ket{{\tt h},N-1-{\tt h}}_A$
\begin{eqnarray}
q_{\tt h} = 1+(N-1-{\tt h}-{\tt h}) = N-2{\tt h}.
\end{eqnarray}
In particular, for the NOON state, ${\tt h}=0$, and therefore
	\begin{eqnarray}
	q_0 =  N,
	\end{eqnarray}
	is the highest possible coherence order for the $N$-qubit STR.  As we shall see later, large coherence orders are exploited in several applications, wherein NOON state offers the maximum advantage.

\subsection{Thermal initial state}
\begin{figure}
	\centering
	\includegraphics[trim=0cm 5.5cm 0cm 0cm,width=5cm,clip=]{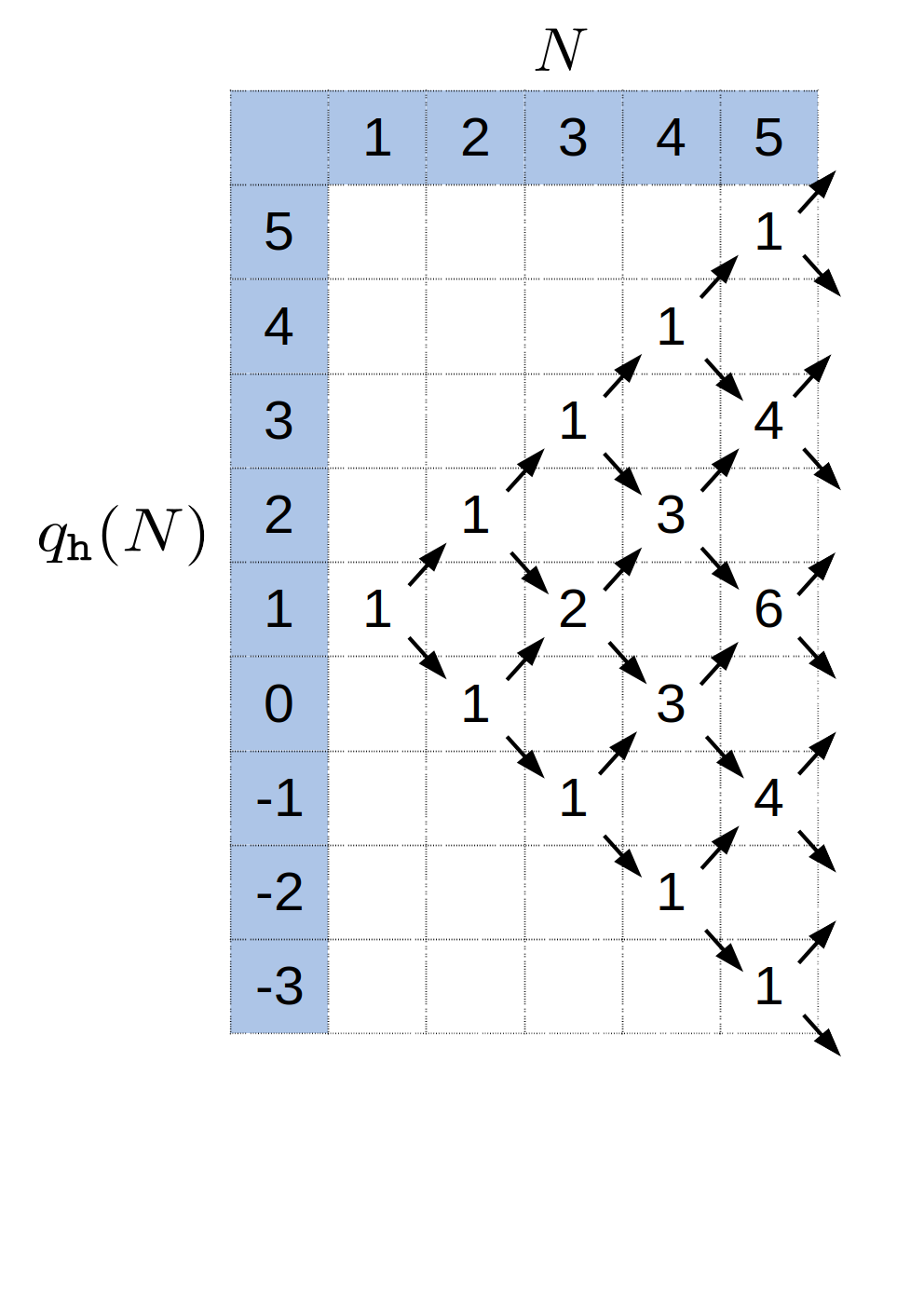}
	\caption{Pascal triangle showing the relative distribution $P(q_{{\tt h}})$ of NOON states (upper most 1's) and MSSM states of various coherence orders $q_{{\tt h}}$ versus size $N$ of the STR.}
	\label{mqpascal}
\end{figure}

Under normal NMR conditions, the initial state of an STR is not the pure ground state considered before, but rather a thermal state as described by Eqn. \ref{starthermalstate}.  We therefore analyze following circuit for the thermal state. \\
\hspace*{3cm}
\Qcircuit @C=1em @R=1em {
	 &  \multigate{1}{\{q_{\tt h}\}} & \qw   \\
	 &  \ghost{\{q_{\tt h}\}} & \qw 
	\inputgroupv{1}{2}{.75em}{.75em}{\rho_{CA}}
} \\
\\
We shall first consider its effect on a particular state vector in the $\ket{0}_C$ subspace, i.e.,
\begin{eqnarray}
&&\ket{0}_C\ket{N-1-{\tt h},{\tt h}}_A
\xrightarrow{H^C, ~CNOT}
\nonumber  \\
&&\frac{1}{\sqrt{2}}\ket{0}_C\ket{N-1-{\tt h},{\tt h}}_A+
\frac{1}{\sqrt{2}}\ket{1}_C\ket{{\tt h},N-1-{\tt h}}_A,~~~
\end{eqnarray}
which is an entangled coherence of coherence order
$q_{{\tt h}} = N-2{\tt h}$.
Note that $q_{{\tt h}} \in\{N,N-2,\cdots, -N+2\}$ for ${{\tt h}}\in\{0,1,\cdots, N-1\}$.  The variants of NOON states with coherence order $q_{{\tt h}} < N$ are known as \textit{many-some, some-many} (MSSM) states, since they are superpositions of state vectors having both 0's and 1's.  Fig. \ref{mqpascal} represents the relative distribution $P(q_{{\tt h}})$ of NOON and MSSM states of various coherence orders in terms of a pascal triangle.

One can also see that the two subspaces lead to identical coherence orders. Thus we may write the output of the above quantum circuit in terms of coherence order as,
\begin{eqnarray}
\rho_{CA} \longrightarrow 
\sum_{q_{{\tt h}}=N,N-2,\cdots}^{-N+2} p_{q_{{\tt h}}} \rho_{q_{{\tt h}}},
\label{mixmssm}
\end{eqnarray}
where $\rho_{q_{{\tt h}}}$ corresponds to mixtures of MSSM states of coherence order $q_{{\tt h}}$ and
\begin{eqnarray}
p_{q_{{\tt h}}} = (p_{0,{{\tt h}}}+p_{1,{{\tt h}}})\frac{P(q_{{\tt h}})}{\sum_{q_{{\tt h}}}P(q_{{\tt h}})}.
\end{eqnarray}

Of course, in principle, it is possible to generate NOON or MSSM states in registers with other topologies too, but not as efficiently as in an STR.  For instance, in a linear chain with nearest-neighbor interaction, the generation of NOON and MSSM states proceeds sequentially as depicted by the arrows in Fig. \ref{mqpascal}.  

In the following, we shall discuss various applications of such quantum correlated states.

\section{Applications}
\subsection{Quantum Sensing of Magnetic Fields 
\label{sec_qs}}
Quantum sensing has a wide range of applications such as rotation sensing (via sagnac interferometric technique) in navigation systems \cite{sagnac_interferprl}, very weak magnetic field sensing, dark matter axion detection \cite{darkmatterdetect_metro}, laser interferometer gravitational wave detection, cold-atom
scanning probe microscope \cite{Atomchipreview}, quantum metrology, nanoscale magnetometry, surface characterization, and molecular spectroscopy \cite{degen2017quantum}.
Here one usually employs the interferometric technique that produces fringes depending on the path difference between two arms of an interferometer. Equivalently, a single spin prepared in a superposition state $\ket{0} + \ket{1}$ is evolved as
\begin{eqnarray}
\ket{0} + \ket{1} \stackrel{\phi}{\longrightarrow}\ket{0} + e^{i\phi} \ket{1},
\end{eqnarray}
 where $\phi$ is the net phase factor due to the magnetic perturbation acting over a certain duration. Typically, one obtains a signal as a function of $\phi$, such as $\expv{I_x} = \cos\phi$ for the above output state. Having $N$ independent spins, instead of one,  improves the signal strength by a factor of $N$.  However, if noise (say, shot-noise) also scales as $\sqrt{N}$, an effective improvement in the signal to noise ratio is only $\sqrt{N}$.  Thus the precision of sensing is limited by the standard deviation of random errors that scales as $1/\sqrt{N}$, which is known as the \textit{standard quantum limit}.  Instead, we now let the magnetic perturbation act on the $N$-particle system prepared in the NOON state. In this case, 
\begin{eqnarray}
\ket{N,0} + \ket{0,N} \stackrel{\phi}{\longrightarrow}\ket{N,0} + e^{i N \phi} \ket{0,N},
\end{eqnarray}
leading to an error scaling of $1/N$, thus achieving the so called \textit{Heisenberg limit}.  In practice, however, the shorter coherence time of the NOON state, compared to the single-spin superposition, poses a challenge in achieving Heisenberg limit.  Jonathan Jones and co-workers have utilized a 10-spin STR in an NMR setup to sensitively measure magnetic fields of the order of a few $\mu T$ and thereby demonstrated beating the standard quantum limit \cite{jones2009magnetic}.

\subsection{Measurement of Translational Diffusion}
Driven by thermal energy, the molecules in a liquid undergo random translational motion that is characterized by the diffusion constant $D$.  NMR has been a powerful tool to measure translational diffusion constant  \cite{price1997pulsed,price1998pulsed}. The standard technique involves letting spins prepared in a superposition state $\ket{0}+\ket{1}$ evolve under a canceling pair of pulsed-field-gradients (PFGs) $G_z$ and $-G_z$ separated by a time delay $\Delta$. 
The PFGs, each of duration $\delta$, introduce  $z$-coordinate dependent phase shifts $\pm \gamma z G_z \delta$, which for a static molecule cancel each other and restore the original superposition state.  
However, a diffusing molecule changes its position after the first gradient and therefore instead of exactly cancelling the gradient-induced phases, it acquires a phase shift, i.e.,
\begin{eqnarray}
\ket{0}+\ket{1} &\stackrel{d_z}{\longrightarrow}&
\ket{0}+e^{i\phi}\ket{1}
~~\mbox{with},
\nonumber \\
\phi &=& \gamma d_z G_z \delta,
\end{eqnarray}
where $d_z$ is the net displacement along the $z$ coordinate during the delay $\Delta$.  In practice, one starts with the thermal equilibrium of the ensemble system $\rho_{eq} = \mathbbm{1}/2 + \epsilon I_z$ and measures a signal that is proportional to the net transverse magnetization $\expv{I_x}$.  
The diffusion experiment can be described by the following quantum circuit.
\\
\\
\vspace*{0.5cm}
\hspace*{1.5cm}
\Qcircuit @C=1em @R=1em {
	& \lstick{\rho_{eq}} &  \gate{H} &   \gate{G_z}  & \gate{\Delta} &  \gate{-G_z} &  \meterB{\expv{I_x}} & \\
} \\
The statistical average of the phase-shifted superpositions lead to a decay of the overall signal $S$ according to
\begin{eqnarray}
S = \exp(-\gamma^2 G_z^2 \delta^2 D \Delta),
\end{eqnarray}
where $D$ is the diffusion constant \cite{price1997pulsed}.  Thus the NMR diffusion experiments involves collecting a series of signals $S$ as a function of the gradient strengths $G_z$ for a fixed diffusion delay $\Delta$.  

If  the diffusing molecule, or a part of it, has an STR, then we can measure diffusion far more efficiently, by exploiting the robust preparation of NOON or MSSM states.  If an STR diffuses along the $z$-coordinate by $d_z$, the NOON/MSSM state would evolve as 
\begin{eqnarray}
\ket{0}_C\ket{N-{\tt h}-1,{\tt h}}_A+\ket{1}_C\ket{{\tt h},N-{\tt h}-1}_A \stackrel{d_z}\longrightarrow
\nonumber \\
\ket{0}_C\ket{N-{\tt h}-1,{\tt h}}_A+e^{i\phi_{{\tt h}}}\ket{1}_C\ket{{\tt h},N-{\tt h}-1}_A,~~~~
\end{eqnarray}
where, 
\begin{eqnarray}
{\tt h} &\in& \{0,1,\cdots,N-1\}
\nonumber \\
\phi_{{\tt h}} &=& l_{\tt h} \gamma_C d_z G_z \delta = l_{{\tt h}}\phi_C~~ \mbox{and,}
\nonumber \\
l_{{\tt h}} &=& 1+(N-2{\tt h}-1)\frac{\gamma_A}{\gamma_C} = 1+(q_{{\tt h}}-1)\frac{\gamma_A}{\gamma_C}.~~
\end{eqnarray}
Here $q_{{\tt h}}$ is the coherence order.
Thus, compared to the diffusion of a single C-spin, phase encoding of diffusion in the STR is improved by the factor $l_{{\tt h}}$ called lopsidedness of the NOON/MSSM state.

The following circuit represents the experimental scheme. \\

\hspace*{0.5cm}
\Qcircuit @C=1em @R=1em {
	\lstick{\rho_C} &  
	\multigate{1}{\{q_{\tt h}\}} &   \multigate{1}{G_z} & \multigate{1}{\Delta} &  \multigate{1}{-G_z} &
	\multigate{1}{q_{\tt h} \Arrow{.2cm} 1} &
	\meterB{\expv{I_x}} \\
	\lstick{\rho_A} &  
	\ghost{\{q_{\tt h}\}} & 
	\ghost{G_z} & 
	\ghost{\Delta} & 
	\ghost{-G_z} & 
	\ghost{q_{\tt h} \Arrow{.2cm} 1}  & 
} \\
\vspace{0.2cm}
\\
Starting from the thermal equilibrium state $\rho_C+\rho_A$ of the STR, one prepares NOON/MSSM states before applying opposite PFGs separated by a diffusion delay.  
Here, a diffusing STR acquires a phase factor by time-separated and mutually canceling pair of gradients.  However, since coherences with order $q_{\tt h} \neq 1$ are not directly observable as transverse magnetization, they have to be converted into observable single-quantum coherence with the help of an untangling CNOT gate.  Another pair of PFGs $g_z$ and $-l_{\tt h}g_z$ help to filter out selectively the signal originating from a specific coherence order $q_{\tt h}$. \\
\\
\hspace*{0.5cm}
\Qcircuit @C=1em @R=1em {    & 
	\multigate{1}{q_{\tt h} \Arrow{.2cm} 1} &
	\qw &
	\push{\rule{-1em}{2em}\equiv} &
	\multigate{1}{g_z} &
	\ctrl{1} & 
	\multigate{1}{-l_{\tt h} g_z} &
	\qw
	\\
	& 
    \ghost{q_{\tt h} \Arrow{.2cm} 1} &
    \qw &
    &
    \ghost{g_z} &
    \targ &
    \ghost{-l_{\tt h} g_z} &
    \qw
   } 
\\

\begin{figure}
	\centering
	\includegraphics[trim=0cm 0cm 0cm 0cm,width=8cm,clip=]{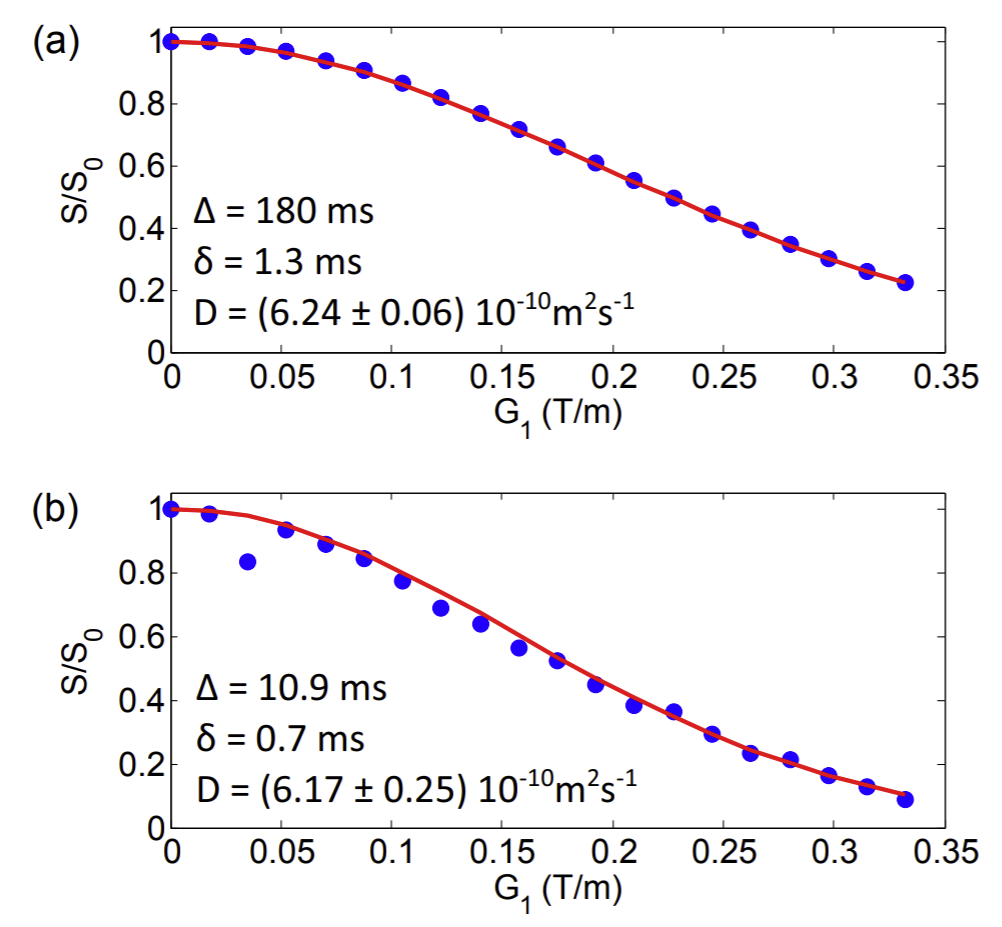}
	\caption{The normalized echo intensities as a function of $G_z$ with (a) standard method and (b) the NOON state method. The dots represent the experimental data and the lines represent the linear fit. (Reused from \cite{shukla2014noon} Copyright (2014) with permission from Elsevier.)}
	\label{noondiff}
\end{figure}

The resulting signal of the central spin decays with PFG strength as
\begin{eqnarray}
S_{\tt h} = \exp(-l_{\tt h}^2 \gamma_C^2 G_z^2 \delta^2 D \Delta).
\end{eqnarray}
The quadratic dependence on the lopsidedness $l_{\tt h}$ drastically improves the diffusion sensitivity of STRs over single-spin systems, particularly for large values of $l_{\tt h}$.  This fact can be exploited to decrease the standard error, or to reduce diffusion time $\Delta$, or to minimize hardware constraints by reducing PFG strength $G_z$, or a combination of these.  While this technique is also known as multiple-quantum diffusion spectroscopy in NMR literature \cite{kay1986application,chapman1993sensitivity}, the STRs provide a unique advantage of parallel and efficient preparation of NOON and MSSM states, and thereby are most suited for this purpose.  Abhishek et al. \cite{shukla2014noon} demonstrated efficient measurement of translation diffusion constant by preparing a 10-qubit NOON state using nine $^1$H spins interacting with a central $^{31}$P spin in trimethyl phosphite (see Fig. \ref{noondiff}).  Compared to the single-qubit experiment, the NOON state experiment required considerably shorter diffusion times to observe the diffusion induced signal decay.  This technique can potentially be adopted for fast measurement of diffusion, for instance in dynamic sample conditions.  One can also utilize NOON states for the characterizing very slow diffusion or to measure diffusion with limited PFG strengths. 

\subsection{RF inhomogeneity mapping}
Sensitive phase encoding by NOON states can open a wide variety of applications, one being in characterizing the spatial amplitude distribution of radio-frequency (RF) pulses used in NMR experiments.  RF inhomogeneity (RFI) depends on several factors such as proximity and the finite size of the RF coils compared to the sample volume, the dielectric constant of the solvent and associated skin-depth, presence of other coils or protective layers, etc.  In quantitative spectroscopy as well as quantum information experiments, it is important to take into account the distribution of RF amplitudes to design high-fidelity quantum controls that generate precise spin dynamics.  The standard method involves recording Rabi oscillations decaying under RFI, also known as Torrey oscillations, whose Fourier transform quantifies the RFI distribution \cite{torrey1949transient, pravia2003robust}.  Since it requires a sufficiently long RF pulse to capture the decay profile, one needs to apply suitably low RF power to avoid overheating of the coil, thereby limiting the RFI measurements to only low powers.  Abhishek et al \cite{shukla2014noon} exploited the sensitive phase encoding of NOON states to a faster mapping of RFI decay profiles, thus allowing its characterization even at higher powers.  Their method can be understood via the following circuit.
\\

\hspace*{0.7cm}
\Qcircuit @C=1em @R=1em {
	\lstick{\rho_C} &  
	\multigate{1}{\{q_{\tt h}\}} & 
	\gate{Y} &
	\gate{X_{\mathrm{Rabi}}} & 
	\gate{-Y} & 
	\multigate{1}{q_{\tt h} \Arrow{.2cm} 1} & 
	\meterB{\expv{I_x}} 
	\\
	\lstick{\rho_A} &  
	\ghost{\{q_{\tt h}\}} & 
	\gate{Y} & 
	\gate{X_{\mathrm{Rabi}}} & 
	\gate{-Y}  & 
	\ghost{q_{\tt h} \Arrow{.2cm} 1} &  
	\qw
} \\
\\

Here the $Y$ gates sandwiching the long $X$ gates that drive Rabi oscillations of the C and A spins allow capturing Torrey oscillations by phase-encoding of NOON/MSSM states.  With this circuit, we can map the two-dimensional profile of 
the probability distribution $P(\Omega_C,\Omega_A)$.  This information can then be incorporated into quantum control sequences to generate robust quantum operations, such as gradient ascent pulse engineering (GRAPE) \cite{khaneja2005optimal}, bang-bang (BB) quantum control \cite{bhole2016steering}, push-pull engineering \cite{innocenti2020ultrafast}, etc.

\subsection{Noise spectroscopy}
The inevitable presence of the surrounding environment leads to the loss of information stored in a quantum register. This phenomenon of decoherence causes destruction of quantum coherence and correlations among qubits, which are essential for robust implementation of quantum technologies in a scalable manner \cite{zurek2003decoherence}. Though noisy devices operating with few qubits, popularly known as \textit{Noisy Intermediate Scale Quantum (NISQ) devices} \cite{preskill2018quantum}, can already perform valuable computational tasks, mitigating decoherence is the key to exploit the full potential of quantum information devices.   In this regard, various quantum control strategies have been developed over the years to combat decoherence \cite{suter2016colloquium}. To ensure maximum efficiency, these control strategies must be judiciously optimized according to the noise characteristics of the surrounding  environment. 
Fortunately, most of the relevant information about the environment is encoded in the noise spectrum.  While sophisticated dynamical decoupling (DD) methods have been developed to suppress noise and combat decoherence \cite{viola1999dynamical}, they also leave behind signatures of frequency-dependent noise profiles on the system dynamics.  In other words, a DD sequence can be designed to filter-in only a particular frequency window of the environmental noise to affect system.
Using this filtering capability, the noise spectroscopy was independently proposed by
Yuge et al. \cite{yuge2011measurement}, and \'Alvarez and Suter \cite{alvarez2011measuring}. Noise spectroscopy procedures with the help of DD sequences are extensively reviewed in  \cite{szankowski2017environmental}.  

In the case of a multi-qubit system, along with the self-noise spectrum of each qubit, knowledge of spatial and temporal correlations among noise fields at various qubits is also important. For example, Sza\'nkowski et al. \cite{szankowski2016spectroscopy} proposed a method to probe spectra of both self and correlated noise between noise sources in two qubits. This method is further generalized to the case of multiple qubits in \cite{paz2017multiqubit}. Multi-qubit noise spectroscopy methods are essentially based on the application of appropriate  DD sequences  after the initialization of the system into various correlated states.  Preparing quantum states with a high degree of correlation in a multi-qubit system is a daunting task in general. However,  as described in section III, the symmetry of STRs allows efficient and robust creation of  specific correlated states (NOON and MSSM states) which makes them particularly suitable to study correlations among noise sources affecting various qubits.  

In Ref. \cite{khurana2016spectral}, Khurana et al. implemented DD based noise spectroscopy protocol with MSSM/NOON states prepared in Trimethyl phosphite (Fig. \ref{starlevels}). The following circuit represents the experimental scheme: \\

\hspace*{1.5cm}
\Qcircuit @C=1em @R=1em {
	\lstick{\rho_C} &  
	\multigate{1}{\{q_{\tt h}\}}  &   \multigate{1}{\mathrm{DD}} & 
	\multigate{1}{q_{\tt h} \Arrow{.2cm} 1}  & 
	\meterB{\expv{I_x}} 
	\\
	\lstick{\rho_A} &  
	\ghost{\{q_{\tt h}\}} & 
	\ghost{\mathrm{DD} }& 
	\ghost{q_{\tt h} \Arrow{.2cm} 1} &
} \\
\\

\begin{figure}
	\includegraphics[trim = 0cm 0cm 0cm 0cm, clip, width=8.5cm ]{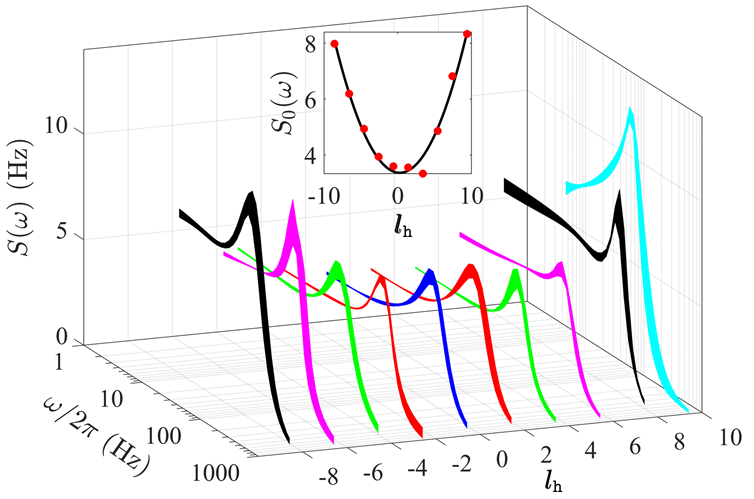} 
	\caption{Trimethylphosphite noise-spectra for various MSSM states with lopsidedness $l_{{\tt h}}$. The dashed lines parallel to $l_{{\tt h}}$-axis represent the maximum frequency (250 Hz) sampled in experiments. The inset shows the scaling of low-frequency spectral density values with $l_{{\tt h}}$. (Reprinted with permission from \cite{khurana2016spectral}, copyright (2016) by the American Physical Society.)}
	\label{MQEns}
\end{figure}

After preparation of NOON or MSSM states, Carr-Purcell-Meiboom-Gill (CPMG) \cite{carr1954effects, meiboom1958modified} DD  sequence is applied on all the qubits, which captures the combined effect of self and correlation noise spectra. The results of the noise spectroscopy of various MSSM states are
shown in Fig. \ref{MQEns}. As expected, the noise spectrum ($S(\omega)$), profiles appear to go higher with the magnitude of the lopsidedness due to an increase in quantum correlations among qubits. Since it is not possible to address ancillary qubits individually in STRs, self-spectra and correlation spectra can not be isolated. However, the variation of combined noise spectrum with lopsidedness provides important information about the spatial correlation among noise sources affecting qubits. For example, in this particular case of Trimethyl phosphite, the  quadratic dependence of low-frequency noise ($S_0(\omega)$) on lopsidedness (inset of Fig. \ref{MQEns}) points to predominantly correlated noise among the qubits \cite{tang1980multiple}. 

\subsection{Algorithmic cooling}
\begin{figure*}
	\centering
	\includegraphics[trim=3.5cm 1.5cm 2.5cm 2cm,width=16cm,clip=]{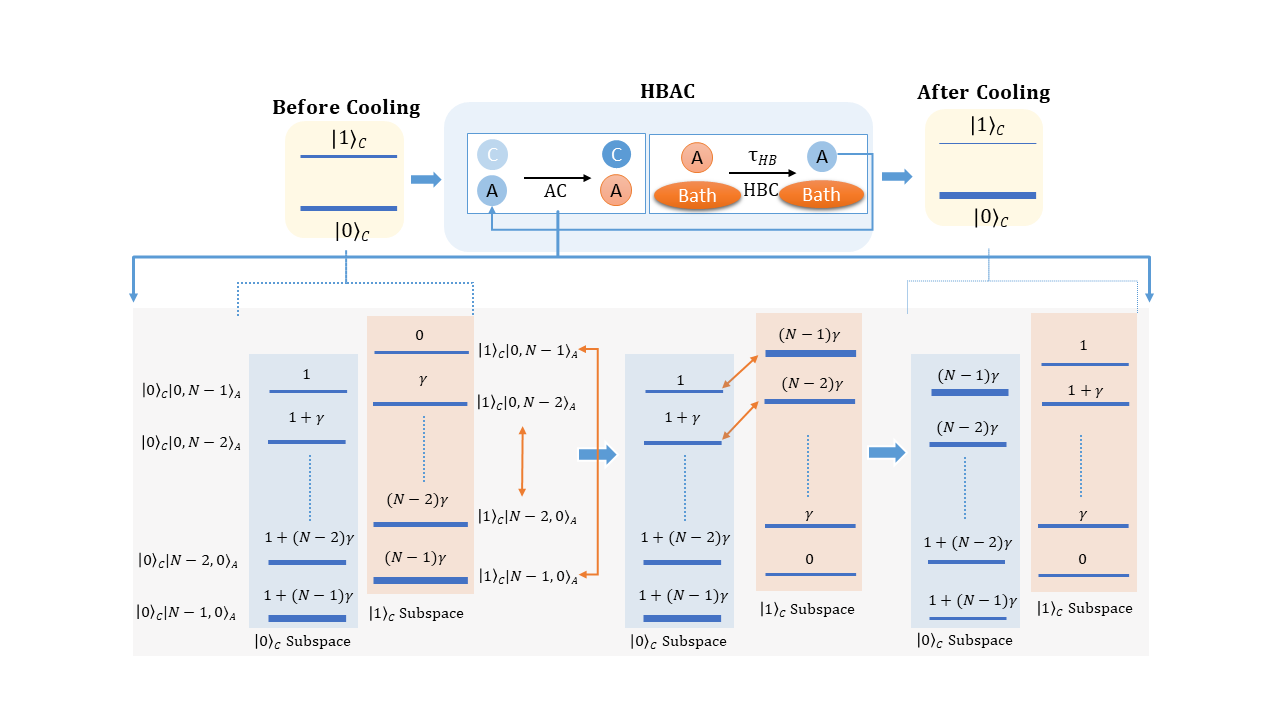}
	\caption{ Schematics of HBAC procedure to purify central qubit using polarization of bunch of surrounding ancillary qubits. Labeling of energy level is as described in Fig. \ref{starlevels}.}
	\label{hbac1}
\end{figure*}
Apart from tackling decoherence, initialization of quantum registers in a high purity quantum state is one of the major requirements for the scalable implementation of fault-tolerant quantum technologies \cite{divincenzo2000physical}.  For quantum processors based on ensembles of identical spin qubits such as NMR and Electron Spin Resonance (ESR), it is extremely challenging to produce pure quantum states in a scalable manner \cite{ladd2010quantum}.  A potential solution to this problem is  Algorithmic Cooling (AC) \cite{sorensen1989polarization, schulman1999molecular}, a protocol that achieves a small set of highly pure quantum bits (\textit{computational} qubits) at the expense of the purity of a large number of ancillary quantum bits (\textit{reset} qubits). Since entropy compression is limited by Shannon bound in a closed system, a non-unitary extension of AC was proposed by Boykin et. al. \cite{boykin2002algorithmic}, which involves iterative entropy removal from the reset qubits to heat bath (i.e. surrounding environment). This protocol is called Heat Bath Algorithmic Cooling (HBAC) and it is extensively reviewed in \cite{park2016heat}. Each iteration of HBAC consists of  two steps: (i) AC, wherein entropy is transferred from the computation qubit to the reset qubit, and (ii) heat-bath cooling (HBC), wherein the reset qubits exchange  the extra
entropy gained to the heat bath.  The efficiency of HBAC critically depends  on the ratio of $T_1$ relaxation of the computational qubit to reset qubits and it should be as large as  possible so that multiple iterations of HBAC can be carried out.

STRs are well-suited quantum registers for AC and HBAC as (i) central spin is connected to a large number of ancillary qubits for strong polarization transfer and (ii) these registers can be efficiently controlled by quantum control methods such as GRAPE and BB optimal control due to their symmetry. AC and HBAC protocols for an STR and corresponding changes in the level-populations are illustrated in Fig. \ref{hbac1}. 

\begin{figure}
	\centering
	\includegraphics[trim=0cm 0cm 0cm 0cm,width=9cm,clip=]{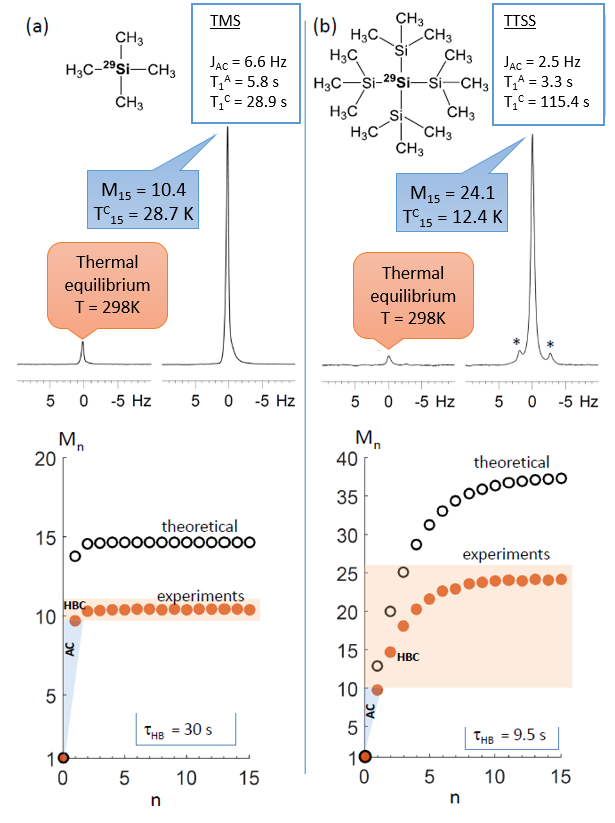}
	\caption{HBAC of (a) TMS and (b) TTSS. Shown in each case
are molecular structure, $^1$H-decoupled $^{29}$Si spectra before and after
HBAC, and magnetization versus HBAC iteration number (n). $M_n = T/ T_n^C$ represent relative magnetization of  central qubit after n HBAC iterations, where $T$ and $T_n^C$, respectively, are spin temperature at thermal equilibrium and after $n$ HBAC iterations separtated by time delay $\tau_{HB}$.   Here the theoretical data points are calculated with ideal HBAC controls.
In the right spectrum, the sidebands (indicated by stars) are due to
$^{29}$Si–$^{13}$C J coupling. (Reprinted  with permission from \cite{pande2017strong}, copyright (2017) by the
American Physical Society.)}
	\label{hbac2}
\end{figure}

Pande et. al. \cite{pande2017strong} conducted a systematic experimental study of HBAC of the central qubit in two STRs: tetramethylsilane (TMS) with $N=13$  and  tetrakis(trimethylsilyl)silane (TTSS) with $N=37$. In these two systems, the $T_1$ relaxation of the central qubit is  longer than that of ancillary qubits by 5 and 34 times respectively. This allowed multiple iterations of HBAC, enhancing the purity of central qubit by 10.4 and 24.1 times, respectively, as shown in Fig. \ref{hbac2}. In TTSS, the strongly boosted polarization of the central qubit allowed the preparation and observation of higher order MSSM states, with coherence orders, as high as $q_{\tt h} = 15$.

\subsection{Amplification of Quantum Fisher Information}
Given a quantum state with an unknown parameter being measured with a suitable observable, the quantum Fisher information (QFI) is a measure of the amount of information that one can extract about the unknown parameter. QFI also quantifies the maximum achievable precision in estimating the unknown parameter with a given amount of resource via quantum Cramer-Rao bound. 
As we discussed in section \ref{sec_qs}, if we do not make use of quantum correlations to estimate/measure an unknown parameter, then we can achieve only the shot-noise scaling in precision. However, by exploiting the quantum correlations we can achieve the Heisenberg scaling and beyond \cite{metro_beyond_heisenberglt}. Here we review the work done in Ref. \cite{CSSK_QFI} which describes  exploiting the ancillary qubits in an STR to amplify QFI as well as to efficiently tomograph the central target qubit.

\textit{Estimation of QFI in an N-qubit STR:}\label{QFI_subsec}
Consider a quantum system prepared in a state in the neighborhood of $\rho_{\theta_0,\phi_0}$ (see Eq. (\ref{rho2})) and
$M$ be a given observable with spectral decomposition $M=\sum_im_i\outpr{m_i}{m_i}$. Let us assume that the polar angle $\theta$ has a distribution around $\theta_0$, while $\phi_0$ is precisely known. Now we may calculate  the probability  $f_{\theta,\phi_0}(m_i)=\mathrm{Tr}(\rho_{\theta,\phi_0}\outpr{m_i}{m_i})$ corresponding to the eigenvalue $m_i$.
Then QFI is defined in terms of non-zero probability  distributions as \cite{Metro_QFI_review}
\begin{eqnarray}
F_\theta(\rho_{\theta_0,\phi_0},M)=\sum_{i}\frac{1}{f_{\theta_0,\phi_0}(m_i)}\left(\at{\frac{\partial f_{\theta,\phi_0}(m_i)}{\partial\theta}}{\theta_0}\right)^2.~~~~~~~
\label{FrhothM}
\end{eqnarray}
where $f_{\theta_0,\phi_0}(m_i)\ne 0$. Here $\at{\partial f_{\theta,\phi_0}(m_i)/\partial\theta}{\theta_0}$ quantifies the sensitivity of the observable $M$ to small fluctuations in $\theta$ around $\theta_0$. An observable which maximises QFI is known as an unbiased observable or symmetric logarithmic derivative (SLD). For example, if the unknown state of a qubit is somewhere near the pole (in Bloch sphere representation), then an observable corresponding to x-component of spin angular momentum (Pauli-x operator) will be the most sensitive one for small fluctuations in $\theta$ (see \cite{CSSK_QFI} for details).

Starting from thermal equilibrium $\rho_{CA}$,
one can utilize the standard NMR technique, namely INEPT \cite{cavanagh,levitt2013spin} to prepare a correlated state of the form
\begin{eqnarray}
\rho_1 = \frac{1}{2^N}\left[\mathbbm{1}_{2^N}+\epsilon_{A} 2 I_{z}^CI_{z}^A\right].
\label{rho1}
\end{eqnarray} 
For a large STR, the above state corresponds to a large anti-phase spin-order  and accordingly leads to a strong NMR signal after applying a suitable read-out pulse. 
Then an unknown state is prepared as follows,
\begin{eqnarray}
\rho_1&& \nonumber \\
\downarrow&& ~e^{-i\theta_0\{\cos(\phi_0+\pi/2) I_{x}^C+\sin(\phi_0+\pi/2) I_{y}^C\}}~~~~~~~ \nonumber \\
\rho_{\theta_0,\phi_0} && = \frac{1}{2^N}\left[\mathbbm{1}_{2^N}+\varepsilon_{A} 2I_{\theta_0,\phi_0}^C I_{z}^A\right].
\label{rho2}
\end{eqnarray} 
The unknown state correlated with the ancillary qubits holds much higher QFI than an uncorrelated qubit.
In general, QFI depends on the size of STR as well as its initial purity.  For high purities, there seems to be little enhancement in QFI.  On the other hand for low purities, like in the case of NMR registers (i.e., $\epsilon_{A}\sim 10^{-5}$), we find empirically that the maximum QFI (which corresponds to SLD) goes as
\begin{eqnarray}
F_\theta(\rho_{\theta_0,\phi_0},M_{\overleftrightarrow{\theta_0},\phi_0}) \approx \epsilon^2_{A}(N-1),
\label{Fthph0}
\end{eqnarray}
where $N \ge 2$.  Sudheer et al demonstrated a QFI amplification of upto 50 times compared to without precorrelation between ancilla and target \cite{CSSK_QFI}. They also showed that quantum discord is the resource responsible for this enhancement in QFI \cite{CSSK_QFI}.  Since QFI is increasing linearly with the size of STR as in Eq. (\ref{Fthph0}), the quantum Cramer-Rao bound for the variance $(\Delta \theta)^2$ in this case is given by
\begin{eqnarray}
(\Delta \theta)^2 \ge \frac{1}{k \epsilon^2_{A}(N-1)}
\end{eqnarray}
where $k$ is the number of identical copies of the unknown state. Hence one can achieve the Heisenberg scaling, if all the copies are correlated, i.e., $N-1=k$.
Finally we note that in NMR, by exploiting the ancillary qubits in an STR, it is possible to know both $\theta_0$ and $\phi_0$ using a single quadrature measurement. However without using the ancillary qubits, the same requires two independent quadrature measurements.  This is akin to ancilla-assisted quantum state tomography \cite{PhysRevA.87.062317}.

\subsection{Temporal ordered phase in star systems}
Motivated in part by high precision 
quantum dynamics achieved in NMR and other experimental platforms, there has been an increasing interest in understanding the physics of quantum systems far out of equilibrium. Decades worth of studies on equilibrium statistical mechanical models have demonstrated a range of ordered phases. A natural question is whether robust ordered phases can exist in quantum systems driven far out of equilibrium. Simplest move away from equilibrium involve a quench \cite{mitraQuench} wherein the Hamiltonian is altered at an instant bringing simple initial states become finite energy density states. Generic many body systems following a quench are expected to eventually reach a state wherein all local quantities approach a thermal ensemble value \cite{AlessioETH,gogolin2016equilibration}. Exceptions are found in integrable systems, where the system approaches generalized Gibbs ensemble \cite{PhysRevB.87.245107,quenchIntegrableAlba,vidmar2016generalized}, and in many body localized systems \cite{MBLquench}. Periodically driven systems further deviate from the equilibrium scenario, and sometimes provide a route to engineer new effective Hamiltonian dynamics and realize phases that cannot exist in equilibrium systems \cite{bruno2013impossibility,PhysRevLett.114.251603}. Analogy between the constant Hamiltonian of equilibrium systems and the constant Floquet unitary of periodically driven systems, however, allow translation of many notions of equilibrium physics such as stationary states, energies etc \cite{Shirley,Sambe}. It has been found that in generic interacting periodically driven quantum systems, irrespective of the initial state, local observables measured at stroboscopic intervals approach a long time steady state described by $T=\infty$ ensemble \cite{ponte2015periodically,AchilleasPRE}. All eigenstates of the Floquet unitary have infinite temperature correlations for local operators. Exceptions can again be found in integrable systems \cite{PhysRevLett.112.150401}, and disordered systems \cite{PhysRevLett.112.150401,lazarides2015fate,abanin2016theory}, with memory of initial conserved quantities or local order persist till long time scales. 

An interesting scenario in periodic driven systems is exemplified by the $Z_2$ symmetric Ising chain kicked with periodic pulses in the transverse direction, whose Floquet unitary is given by 
\begin{eqnarray}
U = \exp\left[-\imath \frac{JT}{\hbar}\sum_{\langle ij\rangle}\sigma^z_i\sigma^z_j\right] \exp\left[\imath \sum_i{h_i}\sigma^x_i \right]\label{Eq:FloquetUnitary}
\end{eqnarray}
where $J$ is the spin interaction strength, $T$ is the time period between successive pulses and $\sigma_i$ are the standard spin operators of the spin $1/2$ representation.
In the vicinity of $h=\pi$, the Floquet eigenstates are symmetric and antisymmetric combinations of short range correlated states {\emph i.e.} of the form $\left | s_1,s_2,s_3\dots \right \rangle\pm \left | -s_1,-s_2,-s_3\dots \right \rangle $, with eigenvalues $\pm \omega$. Here $s_i$ labels the local magnetisation characterising the eigenstate order. It is easy to see that a physical state initialized with $\left | S\right \rangle\equiv\left | s_1,s_2,s_3\dots\right \rangle$ undergoes oscillations between $\left|S\right \rangle$ and $\left | -S\right \rangle$ with period twice that of the drive \cite{KhemaniPhaseStructure}. In the presence of  disorder (say for instance in $h_i$), the eigenstate order and the $\pi$ phase shift of the eigenvalue pairs are stable against additional weak perturbations or finite deviation of mean $h_i$ from $\pi$ \cite{AbsoluteStability,PhysRevLett.118.030401,NayakQTC,PhysRevA.91.033617}. This robust spatio-temporally ordered system is called a discrete time crystal (DTC) and has seen several experimental realizations that match the conditions for DTC up to varying levels of accuracy.\cite{choi2017observation,zhang2017observation,PalDTC,RovnyDTC,AnnRevDTC}

NMR systems \cite{PalDTC,RovnyDTC} form a natural setting for realization of the Floquet unitary described in Eq \ref{Eq:FloquetUnitary}. The STR is a unique system wherein such a Floquet unitary can be realized with a geometry that resembles a mean field version of the the time crystal.
\begin{figure}
\includegraphics[width=0.99\columnwidth]{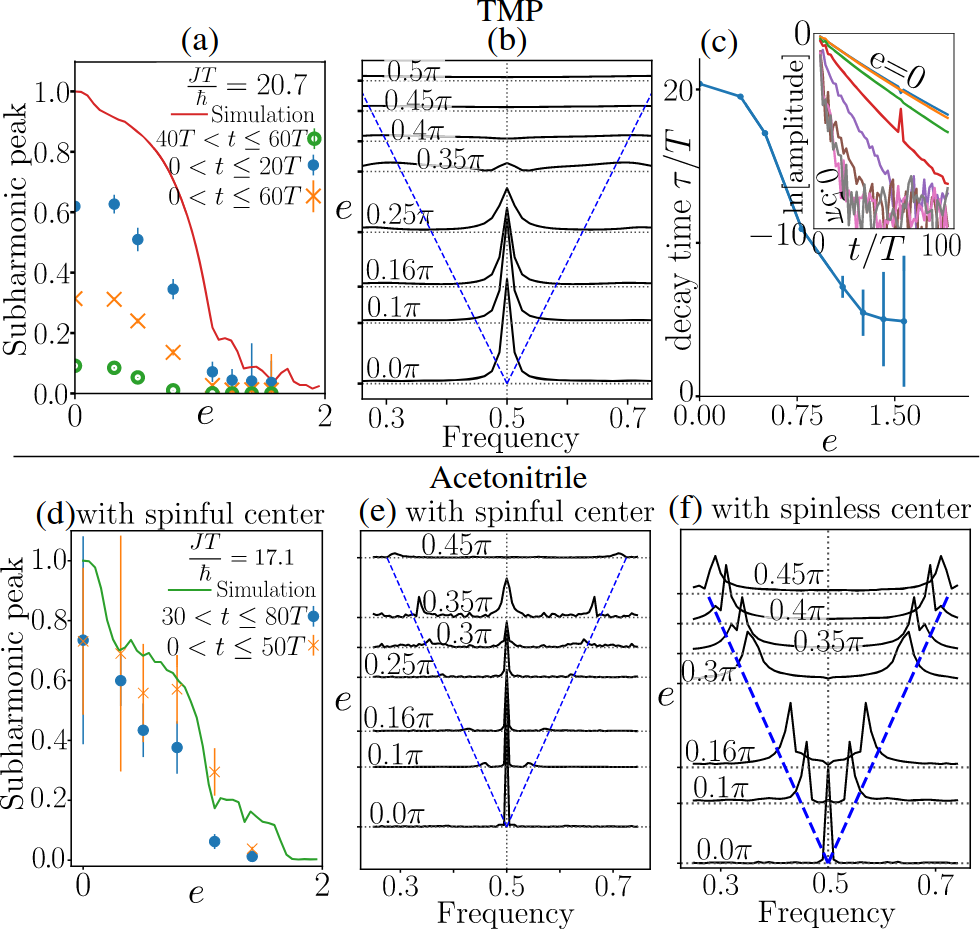}
\caption{ Results of magnetisation measurements on the $10$ spin cluster in Trimethylphosphite (top) and $4$ spin cluster in C-13 Acetonitrile (bottom) subjected to the periodic driving with $h_i\approx \pi-e$. (a) Strength of the subharmonic peak at frequency $0.5/T$ (where $T$ is the time period) in the Fourier power spectrum of the magnetisation as a function of the error $e$ in the $\pi$ pulses. Different markers indicate results for different time windows for Fourier transformations. (b) Waterfall plot of Fourier spectrum of magnetisation time series for different values of $e$ indicating that a stable peak survives upto $e\sim0.25\pi$ as expected in a DTC phase. The blue lines indicate the expected oscillation frequency for an isolated spin.
(c) Decay time scale of the amplitude of the magnetisation oscillations as a function of the error $e$. (d) Similar to panel (a), this shows the strength of the subharmonic peak as a function of the error $e$ in the pulse. (b) Similar to panel (b), Fourier transform of the magnetisation of the C-13 Acetonitrile. (c) Fourier transform of the magnetisation in C-12 acetonitrile which has an NMR inactive central spin indicating a frequency that continuously varies with the error in the applied pulse. (Reprinted  with permission from \cite{PalDTC}, copyright (2018) by the
American Physical Society.)
\label{fig:summaryTemporal}
}
\end{figure}
Pal et al \cite{PalDTC} studied DTC in STRs of sizes up to 37 spins demonstrating robust period two magnetisation response. Fig. \ref{fig:summaryTemporal} shows the summary of the experimental results on the system. A driving protocol identical to Eq. \ref{Eq:FloquetUnitary} was implemented wherein every satellite spin (Ising) interact with the central spin but do not interact with each other. As expected the stable period two oscillations vanish when the central nucleus is replaced with a NMR inactive nucleus. Numerical results showed that for systems initialized with a fixed magnetisation per spin, the time scale over which finite period two oscillations survived increased with number of satellite spins.

Unlike the prototypical example of DTC, the star shaped clusters of spins do not have a extended spatial structure,
neither exhibit a true disorder.  Nevertheless, the system under the action of the Floquet unitary (\ref{Eq:FloquetUnitary}) shows physics that closely resemble the DTC phase with robust subharmonic oscillations even beyond experimentally realistic time scales.

\subsection{Quantum chaos}
Another field where STRs are of potential interest is in quantum chaos. Quantum chaos is the study of quantum systems which, in the classical limit, exhibit chaos. Quantum chaos manifests in interacting spin systems and it is important to understand the fundamental aspects of the phenomenon for developing robust quantum technology, where such interactions might be detrimental. This phenomenon has been extensively studied theoretically \cite{wintgen1986regularity,Haake1991,jensen1992quantum,chaostunneling,stockmann2000quantum,casati2006quantum,bandyopadhyay2004entanglement,chaosspinorbit,chaosSU3}, and investigated experimentally using multiple platforms \cite{hensinger2001dynamical,chaudhury2009quantum,lemos2012experimental,PhysRevA.87.053605,neill2016ergodic,PhysRevE.96.040201,krithika2019nmr}. A natural extension would be the study of large spin systems, which are closer to the classical limit, thus bridging the quantum and classical domains. Star topology systems are ideal candidates for such studies. 

One of the most extensively studied models of chaos is the kicked top model \cite{haake1987classical,ghose2008chaos,lombardi2011entanglement,ruebeck2017entanglement,bhosale2017signatures,dogra2019quantum}.  Krithika et al had \cite{krithika2019nmr} used a two-qubit system to investigate quantum chaos in the kicked top.  Extending this to an STR, one obtains a set of two-qubit kicked-tops, each of which is constituted by an ancillary qubit and the common central qubit.
The Hamiltonian for such a model is given by
\begin{eqnarray}
    H &=& \frac{\pi}{2}(I_x^C+I_x^A)\sum_n \delta(t-n\tau) + \frac{k}{2j}\sum_{i=1}^{N-1}(I_z^C+I_z^{A,i})^2
    \nonumber \\
    &\equiv&
    \frac{\pi}{2}(I_x^C+I_x^A)\sum_n \delta(t-n\tau) + 
    2\pi J_{CA} I_z^CI_z^A,
    \label{Hkt}
\end{eqnarray}
where the linear term is a 90\textsuperscript{o} kick about the x-axis on all spins, $k=2\pi J \tau$ is the chaoticity parameter, and $j=1$ is the combined spin-size of each top. The value of $k$ determines the degree of chaos in the system, with small values indicating little chaos, and large values indicating high degree of chaos in the system.  A numerical study of such a model is described below.

\begin{figure}
	\centering
	\includegraphics[trim=1cm 1cm 0cm 0cm,width=9cm,clip=]{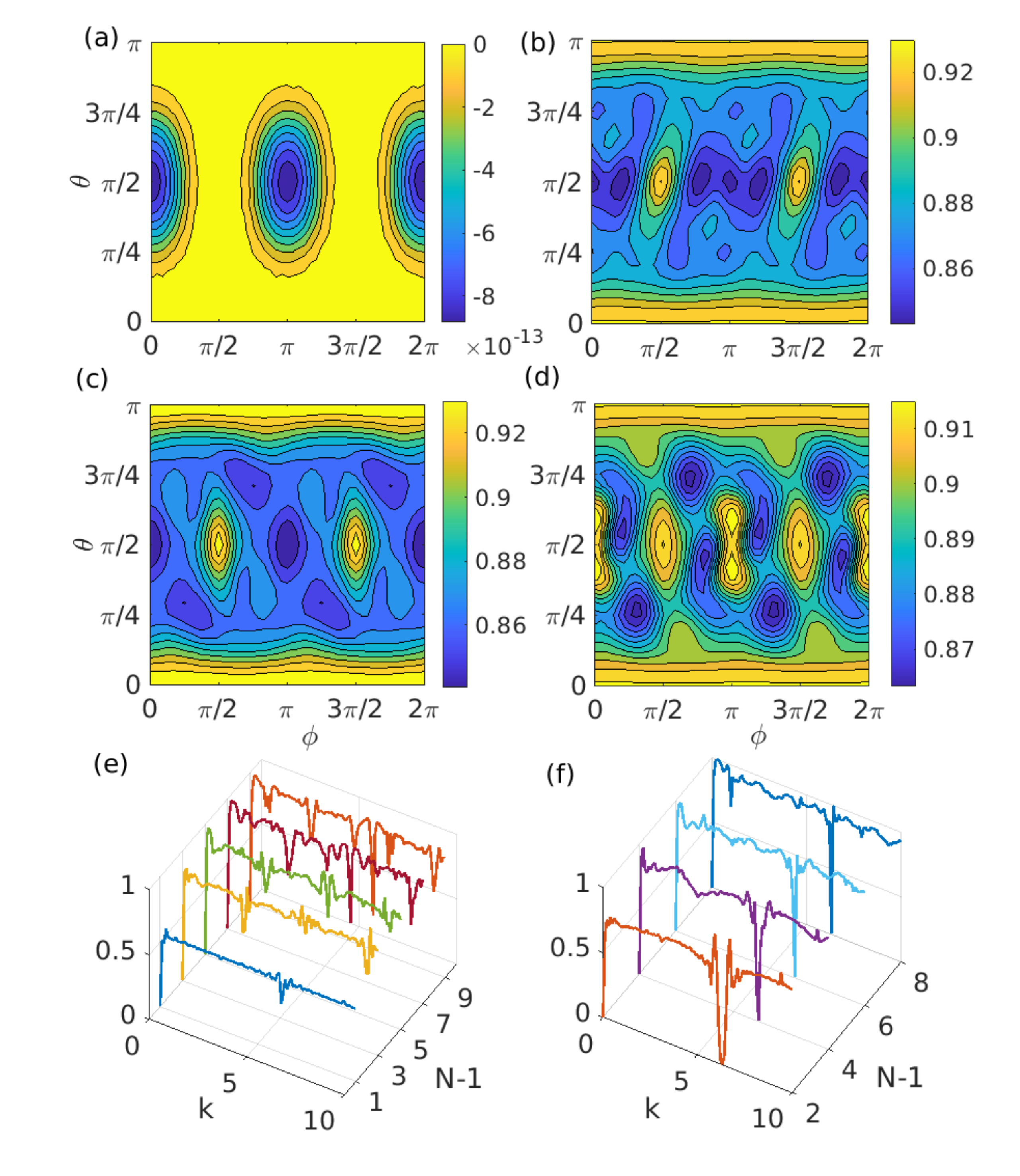}
	\caption{Quantum chaos in an STR.  von Neumann entropy of the central qubit for (a) $k = 0$, (b) $k=2$, (c) $k=5$, and (d) $k=10$. Entropy as a function of STR size for (e) odd and (f) even number of ancillary qubits for the fixed initial state $(\theta,\phi) = (\pi/2,\pi/2)$.
	In all the cases $J_{CA}=50$ Hz and the entropy is averaged over the last 100  out of a total of 200 kicks. 
	}
	\label{phasespace}
\end{figure}

Entanglement entropy has been used as a measure for diagnosing quantum chaos \cite{krithika2019nmr}.  Fig. \ref{phasespace}(a-d) shows the von Neuman entropy of an STR of size $N=10$ with varying chaoticity parameter $k$.  In each case, the STR qubits were initialized to the phase-space point $(\theta,\phi)$ in the Bloch sphere, followed by the evolution under the Hamiltonian in Eq. \ref{Hkt}.
We can see that for $k = 0$ (Fig. \ref{phasespace}(a)), the entropy is identically $\approx 0$ over the entire phase space, and as the chaoticity parameter increases, the average von Neumann entropy of the system increases as well. Moreover, for smaller $k$ values, the phase space shows distinct islands of low entropy surrounded by high entropy regions. Such distributions can be interpreted as a quantum equivalent of regular and chaotic regions of the classical phase space. 

Furthermore, to study the role of STR size in the dynamics of the system, we initialized the system to a fixed point in the phase space $(\theta,\phi) = (\pi/2,\pi/2)$ and evolved the system for 200 kicks while varying the number of ancillas from 1 to 9. The von Neumann entropy of the central spin averaged over the last hundred kicks is shown in Fig. \ref{phasespace} (e)-(f).
Interestingly, we can see that the dynamics shows oscillatory behaviour for both odd and even number of ancillas. For small systems, we can expect such oscillations due to the small dimension of the Hilbert space. However, the system with any even number of ancillas shows more prominent oscillations, compared to odd number of ancillas, in the von Neumann entropy of the central spin for large values of $k$. Such a characteristic can be attributed to symmetries in the system in the different cases considered. Moreover, these results also indicate that the role of multiple-quantum coherences, which is fundamentally linked to the size of the system. The interplay between multiple quantum coherences and chaos is yet to be probed and is crucial for understanding chaos in such systems. 

\subsection{Multi-star systems}
One can find several multi-star topology registers which are interesting for specific applications.  
For example, Pande et al \cite{pande2017strong} have demonstrated purification of two computational central qubits using a double-STR (see Fig. \ref{tops} (e)) wherein each of  computational qubits is coupled  to a separate  set of ancillary qubits.  
Using a pair of central qubits, involving low natural abundance $^{13}$C nuclei, of another double-STR, Khurana et al \cite{khurana2017bang}  demonstrated purification of long-lived singlet states which offer a plethora of applications in spectroscopy as well as imaging \cite{pileio2020long}. In this case, the direct ancilla assisted cooling using the double-star system accelerated the NMR observation of the long-lived singlet state by a factor of 23.  One can envisage multi-star systems wherein individually addressable central qubits form a network amenable for implementing intricate quantum dynamics.

\section{Summary}
Coherent control over the quantum dynamics of large registers is essential to realize quantum supremacy.
A pivotal consideration in designing a large register is the network topology that is optimum for a specific task.  Star-topology network is ideal for parallel implementation of nonlocal quantum gates controlled by a central qubit and acting on all of the indistinguishable ancillary qubits.  The ability to easily and rapidly entangle all qubits is attractive for several applications.  The other advantages of star-registers include the presence of a large ancillary polarization which can be efficiently transferred to the central qubit, spectral simplicity, availability of natural star- or star-like spin systems, etc.
In this review, after introducing the NMR aspects of star-topology systems, we discussed a range of applications demonstrated in some recent works. 
Star topology systems have also been used to extract measures like quantum discord \cite{doronin2015contributions}, out-of-time-order correlations \cite{niknam2020sensitivity} and Renyi entropy \cite{niknam2020experimental} to investigate the dynamics of quantum correlations and processes of information spreading. Hence, star topology systems present a rich test bed for studying multiple facets of quantum dynamics, and should be of interest in most architectures even beyond NMR.

\section*{Acknowledgments}
Funding from DST/SJF/PSA-03/2012-13, CSIR 03(1345)/16/EMR-II, and DST/ICPS/QuST/2019/Q67 is gratefully acknowledged.

\bibliography{revstar}{}

\bibliographystyle{ieeetr}

\end{document}